\documentstyle[prl,aps,twocolumn]{revtex}

\begin{document} 
\draft 
\title{Semiclassical Spectra and Diagonal Matrix Elements by Harmonic 
       Inversion of Cross-Correlated Periodic Orbit Sums}
\author{J\"org Main,$^1$ Kirsten Weibert,$^1$ 
        Vladimir A. Mandelshtam,$^2$ and G\"unter Wunner$^3$}
\address{$^1$Institut f\"ur Theoretische Physik I,
         Ruhr-Universit\"at Bochum, D-44780 Bochum, Germany}
\address{$^2$University California Irvine, Department of Chemistry, 
         Irvine, CA 92612}
\address{$^3$Institut f\"ur Theoretische Physik und Synergetik,
         Universit\"at Stuttgart, D-70550 Stuttgart, Germany}

\date{April 1, 1999}
\maketitle

\begin{abstract}
Semiclassical spectra weighted with products of diagonal matrix elements
of operators $\hat A_\alpha$, i.e., $g_{\alpha\alpha'}(E)=\sum_n 
\langle n|\hat A_\alpha|n\rangle \langle n|\hat A_{\alpha'}|n\rangle /(E-E_n)$ 
are obtained by harmonic inversion of a cross-correlation signal
constructed of classical periodic orbits.
The method provides highly resolved semiclassical spectra even in situations
of nearly degenerate states, and opens the way to reducing the required 
signal lengths to shorter than the Heisenberg time. 
This implies a significant reduction of the number of orbits required for 
periodic orbit quantization by harmonic inversion.

\end{abstract}


The semiclassical quantization of periodic orbits is of fundamental interest 
for both regular \cite{Ber76} and chaotic \cite{Gut67,Gut90} dynamical systems.
In both cases the semiclassical density of states is obtained by summing
over the periodic orbits of the underlying classical system.
However, periodic orbit sums of this type usually diverge.  
Various techniques have been developed in recent years to overcome this 
problem.
Most of these techniques are especially designed for hyperbolic
(chaotic) systems \cite{Cvi89,Aur92,Ber92} and cannot be applied to systems
with regular or mixed regular-chaotic dynamics.
Recently a new technique, based on {\em harmonic inversion} of semiclassical 
signals\cite{Mai97b,Mai98}, was shown to be both very powerful and universal 
for the problem of periodic orbit quantization in that it does not depend on 
special properties of the system such as ergodicity or the existence of a 
symbolic dynamics.
The method only requires the knowledge of periodic orbits and their physical
quantities up to a certain maximum period, which depends on the average local 
density of states.
Unfortunately, this method is not free of the general drawback of
most semiclassical approaches, which suffer from a rapid (exponential for 
chaotic systems) proliferation of periodic orbits with their period, which 
in turn requires an enormous number of orbits to be taken into account.

In this paper we propose to implement an extension of the 
{\it filter-diagonalization method},
which is a method of solving the harmonic inversion 
problem \cite{Wal95,Man97}, to the case of time cross-correlation 
functions \cite{Wal95,Nar97,Man98}, i.e., the one-dimensional time
signal $C(t)$ is extended to a $D\times D$ matrix of cross-correlated 
time signals $C_{\alpha\alpha'}(t)$, with $\alpha,\alpha'=1,\dots,D$.
This method has recently also served as a powerful tool for the 
semiclassical calculation of tunneling splittings \cite{Man98b}.
The informational content of a $D\times D$ time signal is increased roughly 
by a factor of $D$ as compared to a $1\times 1$ signal.
Here we demonstrate that the required amount of periodic
orbits can be significantly reduced when a cross-correlated periodic orbit
sum (signal) is inverted instead of a single periodic orbit sum.
The power of the method is demonstrated for the circle billiard, as an 
example of a completely integrable system.

Consider a quantum Hamiltonian $\hat H$ whose eigenvalues are 
$w_n$ and eigenstates $|n\rangle$. 
We introduce a set of $D$ smooth and linearly independent operators
[there is no need to choose commuting observables]
and define a non-trace type cross-correlated function \cite{Mai99}
\begin{equation}
   g_{\alpha\alpha'}(w)
 = \sum_n {b_{\alpha n}b_{\alpha' n} \over w-w_n+i\epsilon}\ ,
\label{g_ab_qm}
\end{equation}
where $b_{\alpha n}$ and $b_{\alpha' n}$ are the diagonal matrix elements 
of operators $\hat A_\alpha$ and $\hat A_{\alpha'}$, respectively, i.e.\
\begin{equation}
 b_{\alpha n} = \langle n|\hat A_\alpha|n\rangle \; .
\end{equation}
Note that Eq.\ \ref{g_ab_qm} can only be written as a trace formula,
\[
   g_{\alpha\alpha'}(w)
 = \mbox{tr}\left\{\hat A_\alpha\hat G^+(w)\hat A_{\alpha'}\right\}
\]
with the Green Function $\hat G^+(w)=(w-\hat H +i\epsilon)^{-1}$, if either 
$\hat A_\alpha$ or $\hat A_{\alpha'}$ commutes with $\hat H$ \cite{Mai99}.
The weighted density of states is given by
\begin{equation}
   \varrho_{\alpha\alpha'}(w)
 = -{1\over\pi} \, {\rm Im} \, g_{\alpha\alpha'}(w) \; .
\end{equation}
Until recently the periodic orbit quantization would involve
the use of periodic orbit expressions for functions $g_{\alpha\alpha'}(w)$
similar to that in Eq.\ \ref{g_ab_qm}. 
The problem one would then encounter when trying
to extract the semiclassical eigenenergies (poles), is that of the analytic 
continuation of $g_{\alpha\alpha'}^{\rm sc}(w)$ to the real axis, where the
latter diverges.
A general procedure proposed in Refs.\ \cite{Mai97b,Mai98} introduced
aspects of signal processing into the solution of this problem. 
It was suggested to perform a spectral analysis (harmonic inversion) of 
a ``time'' (more precisely, action $s$) signal constructed of the same 
set of periodic orbits.
Here we extend the harmonic inversion procedure of Refs.\ 
\cite{Mai97b,Mai98} to the analysis of a semiclassical cross-correlation 
signal 
\begin{equation}
 C_{\alpha\alpha'}(s) = {1\over 2\pi}\int_{-\infty}^{+\infty}
      g_{\alpha\alpha'}(w)e^{-isw}dw\ ,
\end{equation}
defined as the Fourier transform of $g_{\alpha\alpha'}(w)$. 
When applied to the quantum expression (\ref{g_ab_qm}), this yields
\begin{equation}
C_{\alpha\alpha'}(s)= -i \sum_n b_{\alpha n}b_{\alpha' n} e^{-iw_ns} \; .
\label{C_ab_qm}
\end{equation}
Let us assume that $C_{\alpha\alpha'}(s)$ is given, and the spectral 
parameters (e.g., $w_n$ and $d_{\alpha\alpha',n}=b_{\alpha n}b_{\alpha' n}$) 
are to be extracted. 
This can be done by solving the conventional harmonic inversion problem 
\cite{Wal95,Man97}, which is formulated as a nonlinear fit of the signal 
$C(s)$ by the sum of sinusoidal terms,
\begin{equation}
 C(s)=\sum_n d_ne^{-isw_n}\ , 
\label{Cs}
\end{equation}
with the set of, in general, complex variational parameters $\{w_n,d_n\}$.
Simple information theoretical considerations\cite{Man97,Mai98} then yield an 
estimate for the required signal length, $s_{\rm max}\sim 4\pi\bar\varrho(w)$, 
for poles $w_n\le w$ which can be extracted by this method.
When a periodic orbit approximation of the quantum signal $C(s)$ is used, 
this estimate results sometimes in a very unfavorable scaling because of a 
rapid (exponential for chaotic systems) proliferation of periodic orbits 
with increasing period.

Consider a generalized harmonic inversion problem, which assumes that 
the whole $s$-dependent $D\times D$ signal $C_{\alpha\alpha'}(s)$ is adjusted
simultaneously to the form of Eq.\ \ref{C_ab_qm}, with $b_{\alpha n}$ and 
$w_n$ being the variational parameters.
The advantage of using the cross-correlation approach \cite{Wal95,Nar97,Man98}
is based on the simple argument that the total amount of independent 
information contained in the $D\times D$ signal is $D(D+1)$ multiplied by 
the length of the signal, while the total number of unknowns 
(here $b_{\alpha n}$ and $w_n$) is $(D+1)$ times the total number of poles 
$w_n$. 
Therefore the informational content of the $D\times D$ signal per unknown 
parameter is increased (compared to the case of Eq.\ \ref{Cs}) by a factor 
of $D$. 
[Of course, this scaling holds only approximately and for sufficiently small 
numbers $D$ of operators $\hat A_{\alpha}$ chosen.]

The calculation of a semiclassical approximation to $C_{\alpha\alpha'}(s)$
is significantly simplified for systems with a scaling property, i.e.\
where the shape of periodic orbits does not depend on the scaling 
parameter, $w$, and the classical action scales as 
$S_{\rm po} = w s_{\rm po}$.
For the identity operator $\hat A_1=I$ the element $C_{11}^{\rm sc}(s)$ is 
the Fourier transform of Gutzwiller's trace formula \cite{Gut67,Gut90} for 
chaotic systems, and of the Berry-Tabor formula \cite{Ber76} for regular 
systems, i.e.\ (see Refs.\ \cite{Mai97b,Mai98})
\begin{equation}
   C_{11}^{\rm sc}(s) 
 = \sum_{\rm po} {\cal A}_{\rm po} \delta\left(s-s_{\rm po}\right) \; ,
\label{C_11_sc}
\end{equation}
where $s_{\rm po}$ are the periods of the orbits and ${\cal A}_{\rm po}$
the amplitudes (recurrence strengths) of the periodic orbit contributions 
including phase information.
For $\hat A_1=I$ and an arbitrary smooth operator $\hat A_\alpha$ the elements
$C_{\alpha 1}^{\rm sc}(s)$ are obtained from a semiclassical approximation to 
the generalized trace formula ${\rm tr}\{\hat G^+\hat A_\alpha\}$
\cite{Eck92,Mai98}.
The result is that the amplitudes ${\cal A}_{\rm po}$ in (\ref{C_11_sc})
have to be multiplied by the classical average of the observable $A_\alpha$
along the periodic orbit,
\begin{equation}
 a_{\alpha,{\rm po}} = {1\over s_{\rm po}} \int_0^{s_{\rm po}}
 A_\alpha({\bf q}(s),{\bf p}(s)) ds \; ,
\label{a_po}
\end{equation}
with $A_\alpha({\bf q},{\bf p})$ the Wigner transform of the operator
$\hat A_\alpha$.
The problem of finding a semiclassical approximation to Eq.\ \ref{C_ab_qm}
for the general case of two arbitrary smooth operators $\hat A_\alpha$ and 
$\hat A_{\alpha'}$ is investigated in Ref.\ \cite{Mai99}, where 
numerical evidence is presented
that the amplitudes ${\cal A}_{\rm po}$ in (\ref{C_11_sc}) have to be 
multiplied by the product of the classical averages, 
$a_{\alpha,{\rm po}}a_{\alpha',{\rm po}}$, of the two corresponding
classical observables, i.e.\
\begin{equation}
   C_{\alpha\alpha'}^{\rm sc}(s)
 = \sum_{\rm po} a_{\alpha,{\rm po}} a_{\alpha',{\rm po}} {\cal A}_{\rm po}
   \delta\left(s-s_{\rm po}\right) \; .
\label{C_ab_sc}
\end{equation}
We here adopt the results of Ref.\ \cite{Mai99} and use Eq.\ \ref{C_ab_sc}
as the starting point for the following application of harmonic inversion of 
cross-correlation functions.
Note that all quantities in (\ref{C_ab_sc}) are obtained from the classical
periodic orbits.

The idea of periodic orbit quantization by harmonic inversion 
\cite{Mai97b,Mai98,Nar97} is to fit the semiclassical functions 
$C_{\alpha\alpha'}^{\rm sc}(s)$ given in a finite range $0<s<s_{\rm max}$ 
by the functional form of the quantum expression (\ref{C_ab_qm}).
The frequencies of the harmonic inversion analysis are then identified with
the semiclassical eigenvalues $w_n$ and the amplitudes $b_{\alpha n}$
with the semiclassical approximations to the diagonal matrix 
elements $\langle n|\hat A_\alpha|n\rangle$.
We will show that for a given number of periodic orbits the accuracy of
semiclassical spectra can be significantly improved with the help of the
cross-correlation approach, or, alternatively, spectra with similar accuracy
can be obtained from a periodic orbit cross-correlation signal 
with significantly reduced signal length.

Here we only give a qualitative and brief description of the method.
The details of the numerical procedure of solving the harmonic inversion 
problem (\ref{Cs}) and the generalized harmonic inversion problem 
(\ref{C_ab_qm}) are presented in Refs.\ \cite{Wal95,Man97,Nar97,Man98}.
The idea is to recast the nonlinear fit problem as a linear algebraic
problem \cite{Wal95}. 
This is done by associating the signal $C_{\alpha\alpha'}(s)$
(to be inverted) with a time cross-correlation function between an initial 
state $\Phi_{\alpha}$ and a final state $\Phi_{\alpha'}$,
\begin{equation}
   C_{\alpha\alpha'}(s)
 = \langle\Phi_{\alpha'}|e^{-is\hat H_{\rm eff}}\Phi_{\alpha}\rangle \; ,
\label{ansatz}
\end{equation}
where the fictitious quantum dynamical system is described by an effective 
Hamiltonian $\hat H_{\rm eff}$. 
The latter is defined implicitly by relating its spectrum 
to the set of unknown spectral parameters $w_n$ and $b_{\alpha n}$.
Diagonalization of $\hat H_{\rm eff}$ would yield the
desired $w_n$ and $b_{\alpha n}$. 
This is done by introducing an appropriate basis set in which the matrix 
elements of $\hat H_{\rm eff}$ are available only in terms of the known
signals $C_{\alpha\alpha'}(s)$. 
The Hamiltonian $\hat H_{\rm eff}$ is assumed to be complex symmetric even
in the case of a bound system. This makes the harmonic inversion stable 
with respect to ``noise'' due to the imperfections of the semiclassical 
approximation.
The most efficient numerical and practical implementation of the
harmonic inversion method with all relevant formulas can be found in 
Refs.\ \cite{Man97,Nar97,Man98}.

We now demonstrate the method of harmonic inversion of cross-correlated
periodic orbit sums for the example of the circle billiard.
This is a regular physical system, and the periodic orbits and their physical 
quantities can be obtained analytically.
We choose this system for the sake of simplicity and it will be evident
that the procedure works equally well with more complex systems where
periodic orbits have to be searched numerically.
The nearest neighbor level statistics of integrable systems is a Poisson 
distribution, with a high probability for nearly degenerate states, and we
will demonstrate the power of our new method by fully resolving those nearly 
degenerate states.
The exact quantum mechanical eigenvalues $E=\hbar^2k^2/2M$ of the circle
billiard are given as zeros of Bessel functions $J_{|m|}(kR)=0$, where $m$ 
is the angular momentum quantum number and $R$, the radius of the circle.
In the following we choose $R=1$.
Semiclassical eigenvalues can be obtained by an EBK torus quantization
resulting in the quantization condition \cite{Boa94}
\begin{equation}
 kR\sqrt{1-(m/kR)^2} - |m|\arccos{|m|\over kR} = \pi\left(n+{3\over 4}\right)
\label{EBK}
\end{equation}
with $m=0,\pm 1,\pm 2,\dots$ being the angular momentum quantum number and 
$n=0,1,2,\dots$ the radial quantum number.
States with angular momentum quantum number $m\ne 0$ are twofold degenerate.

For billiard systems the scaling parameter is the absolute value of the
wave vector, $w\equiv k=|{\bf p}|/\hbar$, and the action is proportional to 
the length of the orbit, $S_{\rm po}=\hbar k\ell_{\rm po}$.
The periodic orbits of the circle billiard are those orbits for which the 
angle between two bounces is a rational multiple of $2\pi$, i.e., the periods 
$\ell_{\rm po}$ are obtained from the condition
\begin{equation}
 \ell_{\rm po} = 2m_r \sin \gamma \; ,
\end{equation}
with $\gamma\equiv\pi m_\phi/m_r$, $m_\phi=1,2,\dots$ the number of turns 
of the orbit around the origin, and $m_r=2m_\phi,2m_\phi+1,\dots$ the number 
of reflections at the boundary of the circle.
Periodic orbits with $m_r\ne 2m_\phi$ can be traversed in two directions
and thus have multiplicity 2.
To construct a periodic orbit cross-correlation signal
$C_{\alpha\alpha'}^{\rm sc}(\ell)$, as defined by Eq.\ \ref{C_ab_sc}, we choose 
three different operators, $\hat A_1=I$ the identity, $\hat A_2=r$ the 
distance from the origin, and $\hat A_3=(L/k)^2$ the square of the scaled 
angular momentum.
For these operators the classical weights $a_{\alpha,{\rm po}}$ 
(Eq.\ \ref{a_po}) are obtained as
\begin{eqnarray}
 a_{1,{\rm po}} &=& 1 \nonumber \\
\label{a_circ}
 a_{2,{\rm po}} &=& {1\over 2}\left(1+{\cos\gamma\over 
                    \tan\gamma} {\rm arsinh} \, \tan\gamma\right) \\
 a_{3,{\rm po}} &=& \cos^2\gamma \; . \nonumber
\end{eqnarray}
The calculation of the weights ${\cal A}_{\rm po}$ in Eq.\ \ref{C_ab_sc}
depends on whether the classical dynamics is regular or chaotic.
For the circle billiard with regular dynamics we start from the Berry-Tabor 
formula \cite{Ber76} and obtain
\begin{equation}
   {\cal A}_{\rm po}
 = \sqrt{\pi\over 2}{\ell_{\rm po}^{3/2}\over m_r^2}
   e^{-i({\pi\over 2}\mu_{\rm po}+{\pi\over 4})}  \; ,
\label{A_circ}
\end{equation}
where $\mu_{\rm po}=3m_r$ is the Maslov index.
Note that the formalism is directly applicable to chaotic systems with the
amplitudes ${\cal A}_{\rm po}$ in (\ref{C_ab_sc}) computed 
according to Gutzwiller's trace formula \cite{Gut67,Gut90}.

\begin{figure}[t]
\newpage
\phantom{}
\vspace{10.9cm}
\includegraphics{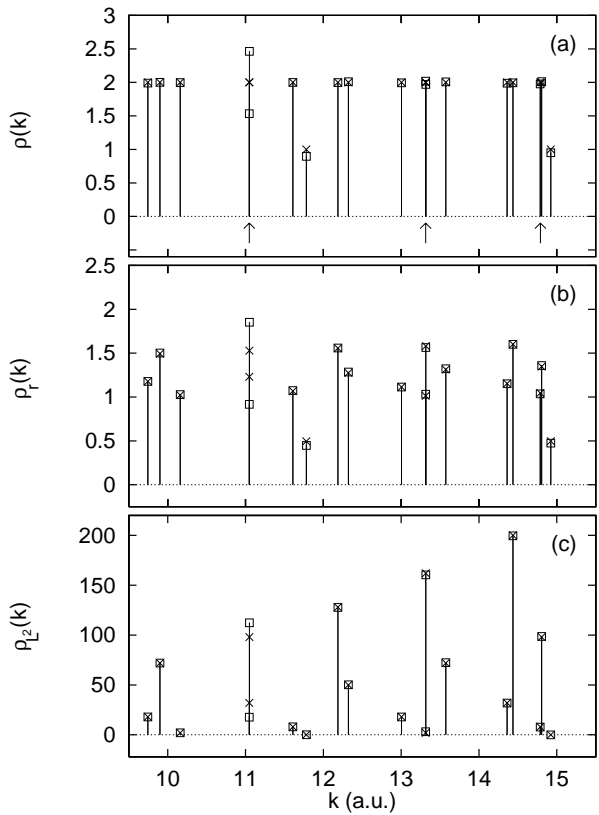}
\caption{\label{fig1} 
Density of states weighted with the diagonal matrix elements of the
operators (a) $\hat A=I$, (b) $\hat A=r$, (c) $\hat A=L^2$ for the
circle billiard with radius $R=1$ as functions of the wave number
$k=|{\bf p}|/\hbar$ (in atomic units).
Crosses: EBK eigenvalues and quantum matrix elements.
Squares: Eigenvalues and matrix elements obtained by harmonic inversion
of cross-correlated periodic orbit sums.
Three nearly degenerate states are marked by arrows.
}
\end{figure}

Once all the ingredients of Eq.\ \ref{C_ab_sc} for the circle billiard are 
available, the $3\times 3$ periodic orbit cross-correlation signal 
$C_{\alpha\alpha'}^{\rm sc}(\ell)$ can easily be constructed and
inverted by the generalized filter-diagonalization method.
Results obtained from the periodic orbits with maximum length $s_{\rm max}=100$
are presented in Fig.\ \ref{fig1}.
Fig.\ \ref{fig1}a is part of the density of states, $\varrho(k)$,
Figs.\ \ref{fig1}b and \ref{fig1}c are the density of states weighted with 
the diagonal matrix elements of the operators $\hat A=r$ and $\hat A=L^2$, 
respectively.
The crosses are the results from the harmonic inversion of the periodic
orbit cross-correlation signals.
For comparison the squares mark the matrix elements obtained by exact
quantum calculations at positions $k^{\rm EBK}$ obtained from the EBK
quantization condition (\ref{EBK}).
We do not compare with the exact zeros of the Bessel functions because
Eq.\ \ref{C_ab_sc} is correct only to first order in $\hbar$ and thus
the harmonic inversion of $C_{\alpha\alpha'}^{\rm sc}(s)$ cannot provide
the exact quantum mechanical eigenvalues.
A discussion of the semiclassical accuracy can be found in \cite{Boa94},
and higher order $\hbar$ corrections to the periodic orbit sum are 
considered in \cite{Mai98b}.
However, the perfect agreement between the eigenvalues $k^{\rm HI}$
obtained by harmonic inversion and the EBK eigenvalues $k^{\rm EBK}$ is
remarkable, and this is even true for nearly degenerate states marked by 
arrows in Fig.\ \ref{fig1}a.
The eigenvalues of some nearly degenerate states are presented in Table
\ref{table1}.
It is important to emphasize that these states with level splittings of, e.g.,
$\Delta k=6\times 10^{-4}$ cannot be resolved by the originally proposed 
method of periodic orbit quantization by harmonic inversion \cite{Mai97b,Mai98}
with a periodic orbit signal length $s_{\rm max}=100$.
To resolve the two levels at $k\approx 11.049$ (see Table \ref{table1})
a signal length of at least $s_{\rm max}\approx 500$ is required if a single 
periodic orbit function $C^{\rm sc}(s)$ is used instead of a cross-correlation
function.
The method presented in this paper can therefore be used to significantly
reduce the required signal length and thus the required number of periodic
orbits 
\begin{table}[t]
\caption{\label{table1}
Nearly degenerate semiclassical states of the circle billiard.
$k^{\rm EBK}$: Results from EBK-quantization.
$k^{\rm HI}$: Eigenvalues obtained by harmonic inversion of cross-correlated
periodic orbit sums. States are labeled by the radial and angular momentum
quantum numbers $(n,m)$.}

\medskip
\begin{center}
\begin{tabular}[t]{rrrr}
  \multicolumn{1}{c}{$n$} &
  \multicolumn{1}{c}{$m$} &
  \multicolumn{1}{c}{$k^{\rm EBK}$} &
  \multicolumn{1}{c}{$k^{\rm HI}$} \\ 
\hline
   1 &  4 &  11.048664 &   11.048569 \\
   0 &  7 &  11.049268 &   11.049239 \\
\hline
   3 &  1 &  13.314197 &   13.314216 \\
   0 &  9 &  13.315852 &   13.315869 \\
\hline
   3 &  2 &  14.787105 &   14.787036 \\
   1 &  7 &  14.805435 &   14.805345 \\
\hline
   1 & 11 &  19.599795 &   19.599863 \\
   5 &  1 &  19.609451 &   19.608981 \\
\hline
   1 & 15 &  24.252501 &   24.252721 \\
   6 &  2 &  24.264873 &   24.264887 \\
\end{tabular}
\end{center}
\end{table}
\noindent
for periodic orbit quantization by harmonic inversion.
As such the part of the spectrum shown in Fig.\ \ref{fig1} can even
be resolved, despite the splittings of the nearly degenerate states marked 
by the arrows, from a short cross-correlation signal with $s_{\rm max}=30$, 
which is about the Heisenberg period $s_H=2\pi\bar\varrho(k)$, i.e.\
half of the signal length required for the harmonic inversion
of a $1\times 1$ signal \cite{Mai98}.
The reduction of the signal length is especially 
important if the periodic 
orbit parameters are not given analytically, as in our example of the circle 
billiard, but must be obtained from a numerical periodic orbit search.
Note also that the density of periodic orbits increases with the period and 
in chaotic systems the proliferation of periodic orbits is exponential.
How small can $s_{\rm max}$ get as one uses more and more operators in the
method?
It might be that half of the Heisenberg period is a fundamental barrier
for bound systems with chaotic dynamics in analogy to the Riemann-Siegel
formula \cite{Ber92} while for regular systems an even further reduction
of the signal length should in principle be possible.
However, further investigations are necessary to clarify this point.

In conclusion, we have introduced a method of periodic orbit quantization
and calculation of diagonal matrix elements based on the construction 
of a cross-correlated periodic orbit sum followed by its harmonic inversion.
The method is not restricted to bound regular systems but is universal
and can be applied to open and chaotic systems as well.
It opens the way to reducing the required signal lengths to shorter than 
the Heisenberg time and therefore significantly reduces the number of 
orbits required for periodic orbit quantization by harmonic inversion.

\bigskip
JM thanks F. Steiner for stimulating discussions.
JM and GW acknowledge the support by the Son\-der\-for\-schungs\-be\-reich 
No.\ 237 of the Deutsche For\-schungs\-ge\-mein\-schaft.
JM is grateful to Deutsche For\-schungs\-ge\-mein\-schaft for a
Habilitandenstipendium (Grant No.\ Ma 1639/3).


\begin{references}
\bibitem{Ber76}  M. V. Berry and M. Tabor, Proc. R. Soc. London,
                 Ser. A {\bf 349}, 101 (1976).
\bibitem{Gut67}  M.\ C.\ Gutzwiller, J. Math. Phys. {\bf 8}, 1979 (1967);
                 {\bf 12}, 343 (1971).
\bibitem{Gut90}  M. C. Gutzwiller, {\it Chaos in Classical and Quantum
                 Mechanics} (Springer, New York, 1990).
\bibitem{Cvi89}  P. Cvitanovi\'c and B. Eckhardt, Phys. Rev. Lett. {\bf 63},
                 823 (1989).
\bibitem{Aur92}  R. Aurich, C. Matthies, M. Sieber, and F. Steiner, 
                 Phys. Rev. Lett. {\bf 68}, 1629 (1992).
\bibitem{Ber92}  M. V. Berry and J. P. Keating, Proc. R. Soc. London,
                 Ser. A {\bf 437}, 151 (1992).
\bibitem{Mai97b} J. Main, V. A. Mandelshtam, and H. S. Taylor, Phys. Rev.
                 Lett. {\bf 79}, 825 (1997).
\bibitem{Mai98}  J. Main, V. A. Mandelshtam, G. Wunner, and H. S. Taylor,
                 Nonlinearity {\bf 11}, 1015 (1998).
\bibitem{Wal95}  M. R. Wall and D. Neuhauser, J. Chem. Phys. {\bf 102},
                 8011 (1995).
\bibitem{Man97}  V. A. Mandelshtam and H. S. Taylor, Phys. Rev. Lett.
                 {\bf 78}, 3274 (1997) and J. Chem. Phys. {\bf 107}, 6756
                 (1997).
\bibitem{Nar97}  E. Narevicius, D. Neuhauser, H. J. Korsch, and N. Moiseyev, 
                 Chem. Phys. Lett. {\bf 276}, 250 (1997).
\bibitem{Man98}  V. A. Mandelshtam, J. Chem. Phys. {\bf 108}, 9999 (1998).
\bibitem{Man98b} V. A. Mandelshtam and M. Ovchinnikov, J. Chem. Phys.
                 {\bf 108}, 9206 (1998).
\bibitem{Mai99}  J. Main and G. Wunner, Phys. Rev. E (1999), submitted.
\bibitem{Eck92}  M. Wilkinson, J. Phys. A {\bf 21}, 1173 (1988);
                 B. Eckhardt, S. Fishman, K. M\"uller, and D. Wintgen,
                 Phys. Rev. A {\bf 45}, 3531 (1992).
\bibitem{Boa94}  T. Prosen and M. Robnik, J. Phys. A {\bf 26}, L37 (1993);
                 P. A. Boasman, Nonlinearity {\bf 7}, 485 (1994).
\bibitem{Mai98b} J. Main, K. Weibert, and G. Wunner, Phys. Rev. E {\bf 58},
                 4436 (1998).
\end{references}
\end{document}